# A new type of boundary layers, a representative model, and surface wrinkling in inhomogeneous plates

A. G. Kolpakov (algk@ngs.ru) and S. I. Rakin (rakinsi@ngs.ru)

"SysAn", A. Nevskogo Street, 34 - 12a, 630075 Novosibirsk, Russia

**Abstract**. We find a new type of boundary layers that occur at the top and bottom surfaces of an inhomogeneous plate. The boundary layers of this type never occur in homogeneous plates or plates made of layers of homogeneous materials. The thickness of such a boundary layer is usually less than that of a one single fiber of the plate.

Moreover, we introduce a new notion of representative plate as a plate whose stress-strain state *in some parts* is similar to the stress-strain state in the original plate. In addition, every part of the original plate has a corresponding part in the representative plate. We demonstrate that a three-layered plate can be representative of a plate formed of an arbitrary number of layers.

We also demonstrate a wrinkling phenomenon that happens on the top and bottom surfaces of inhomogeneous plates. This effect does not occur in homogeneous plates or plates made of layers of homogeneous materials. Such wrinkling can become a significant adverse effect, and thus should be given utmost attention.

**1. A new type of boundary layers in periodic plates**. Let us consider a plate of periodic structure formed by repeating a small periodicity cell (PC) $\varepsilon P$ in $Ox_1x_2$-plane, where $\varepsilon \ll 1$ is the characteristic size of the cell (it is also the characteristic thickness of the plate). One can find numerous examples of such plates in everyday life and in industrial production. A typical example is that of fiber-reinforced plates, see Fig.1*a*. An effective tool for analysis of periodic structures is homogenization theory [1-5].

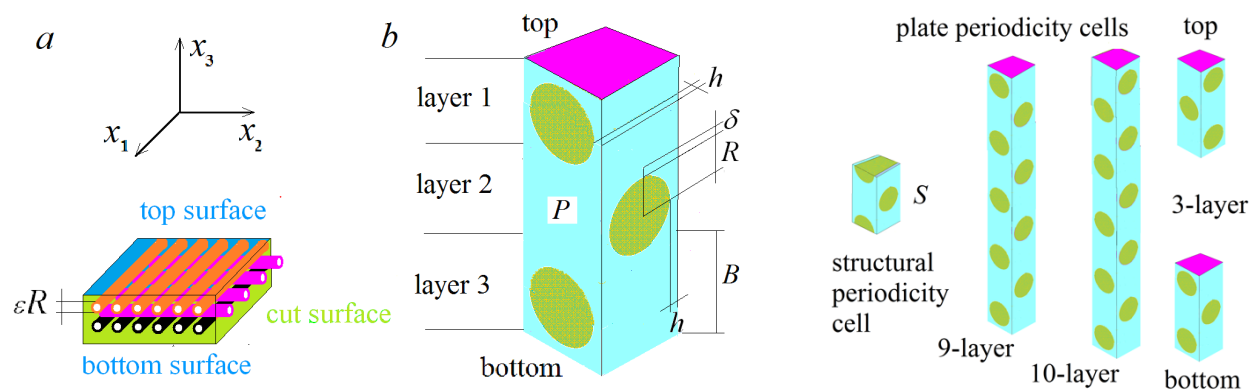

**Fig. 1** *Plate reinforced with fiber layers:*

*the "slow" variables* **x** *- (a); PCs in the "fast" variables* **y** *- (b)*

Homogenization theory on its own does not provide any computational formulas: it reduces all computation to solving a so-called periodicity cell problem (PCP). One specific feature of plates is that their periodicity cells (PCs) extend from top to bottom of the plate and the top and bottom surfaces are free surfaces [4,5] see Fig.1 *b*, *c*. This feature does not appear in solid composites [1-3].

In the literature on plates and shells, boundary layers are associated with transversal cut surfaces, Fig.1*a*. For a classical uniform plate, such kind of boundary layers were intensively discussed by Rayleigh, Love, Lamb, and Basset [6]. In the 1970s-1980s, boundary layers at transversal cut surface were intensively discussed for laminated composites, see e.g. [7, 8]. When considering inhomogeneous plates of general form, we meet a new type of boundary layers associated with the top and bottom surfaces of the plate. Boundary layers of this type never occur in homogeneous plates or in plates made of layers of homogeneous materials. Probably this is the reason why we find no discussion of this kind of boundary layers in the literature. Below, we investigate this new type of boundary layers.

**2. Homogenization method as applied to inhomogeneous plates of periodic structure.** We assume that the PC is given as $P = [0, h_1] \times [0, h_2] \times [-h, h]$ in the "fast" variables $\mathbf{y} = \mathbf{x}/\varepsilon$. This means that the top and bottom surfaces of the plate are flat (it is not necessary so, but it is a convenient assumption for our analysis). Let $\Gamma_i = \{\mathbf{y} : y_i = 0\}$ and $\Gamma_i + h_i \mathbf{e}_i$ denote the opposite faces of the PC $P$. Let $P_2 = [0, h_1] \times [0, h_2]$ denote the projection of the PC $P$ to $Oy_1 y_2$-plane.

We use homogenization theory originally applied to plates in [4,5] and later discussed in numerous publications (see [9] for further references). The solution to the general elasticity theory problem for a periodic thin plate (within homogenization theory) is sought in the form [9]

$$\mathbf{u}^\varepsilon = \mathbf{u}_0(x_1, x_2) + \varepsilon \mathbf{u}_1(x_1, x_2, \mathbf{x}/\varepsilon) + ..., \quad (1)$$

In (1), $\mathbf{y} = \mathbf{x}/\varepsilon$ are the "fast" (microscopic) and $(x_1, x_2)$ are the "slow" (macroscopic) variables [4,5], $\mathbf{x} = (x_1, x_2, x_3)$.

In (1), $\mathbf{u}_0(x_1, x_2) = (u_{01}, u_{02}, u_{03})(x_1, x_2)$ denotes the solution to the homogenized problem ($u_{01}$ and $u_{02}$ are the in-plane displacements, and $u_{03}$ is the normal deflection); $\varepsilon \mathbf{u}_1(x_1, x_2, \mathbf{x}/\varepsilon)$ is the corrector.

The corrector $\varepsilon \mathbf{u}_1(x_1, x_2, \mathbf{x}/\varepsilon)$ is small, but its derivatives (thus, its contribution to the local SSS) can be large. The corrector has the form $\varepsilon[e_{\alpha\beta}(x_1, x_2)\mathbf{N}^{\alpha\beta 0}(\mathbf{x}/\varepsilon) + \rho_{\alpha\beta}(x_1, x_2)\mathbf{N}^{\alpha\beta 1}(\mathbf{x}/\varepsilon)]$ [4,5], where $\varepsilon^0_{\alpha\beta} = \partial u_{0\alpha}/\partial x_\beta$ are the macroscopic in-plane tensions/shift, $\rho_{\alpha\beta} = \varepsilon^1_{\alpha\beta} = \partial u_{03}/\partial x_\alpha \partial x_\beta$ are the macroscopic curvatures/torsion, and $\mathbf{N}^{\alpha\beta\nu}(\mathbf{y})$ is the solution to the following PCP:

$$\begin{cases} (a_{ijkl}(\mathbf{y})N^{\alpha\beta\nu}_{k,ly} + (-1)a_{ij\alpha\beta}y^{\nu}_3)_{,jy} = 0 \text{ in } P, \\ (a_{ijkl}(\mathbf{y})N^{\alpha\beta\nu}_{k,ly} + (-1)a_{ij\alpha\beta}y^{\nu}_3)n_j = 0 \text{ at } y_3 = \pm h, \\ [\mathbf{N}^{\alpha\beta\nu}(\mathbf{y})]_\alpha = 0 \quad (\alpha=1,2) \end{cases} \qquad (2)$$

Hereafter, $a_{ijkl}(\mathbf{y}) = a^F_{ijkl}$ in the fiber and $a_{ijkl}(\mathbf{y}) = a^M_{ijkl}$ in the matrix, where $a^F_{ijkl}$ and $a^M_{ijkl}$ are the elastic constants of the fibers and the binder, correspondingly. The indices $\nu$ and $\mu$ take values 0, 1; the other Greek indices take values 1, 2; Latin indices take values 1, 2, 3; $[f(\mathbf{y})]_\alpha$ means the "jump" of the function $f(\mathbf{y})$ on the opposite faces $\mathbf{y} \in \Gamma_\alpha$ and $\mathbf{y} \in \Gamma_\alpha + h_\alpha \mathbf{e}_\alpha$ of the PC $P$. The equality $[f(\mathbf{y})]_\alpha = 0$ implies periodicity of $f(\mathbf{y})$ in $y_\alpha$.

For $\alpha\beta = 11$ and 22, PCP (2) corresponds to the macroscopic tension ($\nu = 0$) or bending ($\nu = 1$). For $\alpha\beta = 23$, it corresponds to the macroscopic shifts ($\nu = 0$) or torsion ($\nu = 1$).

The equation $a_{ijkl}(\mathbf{y})\xi^{\alpha\beta\nu}_{k,l} = y^{\nu}_3 a_{ij\alpha\beta}(\mathbf{y})$ is solvable with respect to the function $\boldsymbol{\xi}^{\alpha\beta\nu}(\mathbf{y})$ for any $\alpha, \beta, \nu$. The explicit formulas for $\boldsymbol{\xi}^{\alpha\beta\nu}(\mathbf{y})$ may be found in [10]. Let us define $\mathbf{Z}^{\alpha\beta\nu}(\mathbf{y}) = \varepsilon^{\nu}_{\alpha\beta}(\mathbf{x})[\mathbf{N}^{\alpha\beta\nu}(\mathbf{y}) + \boldsymbol{\xi}^{\alpha\beta\nu}(\mathbf{y})]$, and thus obtain from (1) the following problem:

$$\begin{cases} (a_{ijkl}(\mathbf{y})Z^{\alpha\beta\nu}_{k,ly})_{,jy} = 0 \text{ in } P, \\ a_{ijkl}(\mathbf{y})Z^{\alpha\beta\nu}_{k,ly}n_j = 0 \text{ at } y_3 = \pm h, \\ [\mathbf{Z}^{\alpha\beta\nu}(\mathbf{y})]_i = \varepsilon^{\nu}_{\alpha\beta}(\mathbf{x})[\xi^{\alpha\beta\nu}(\mathbf{y})]_i \end{cases} \qquad (3)$$

Above, PCP (3) is written in the "fast" variables $\mathbf{y}$, where $\varepsilon^{\nu}_{\alpha\beta}(\mathbf{x})$ are parameters. Any two solutions to PCP (3) differ one from another by rigid body displacements. We eliminate the rigid body displacements by fixing some points of the PC. Note that PCP (3) describes the microscopic deformation of the PC, corresponding to macroscopic in-plane strains $\varepsilon^0_{\alpha\beta}$ and macroscopic curvatures / torsion strains $\rho_{\alpha\beta}$. The local stresses that arise in the PC are computed as

$$\sigma^{\alpha\beta\nu}_{pq}(\mathbf{y}) = a_{pqkl}(\mathbf{y})Z^{\alpha\beta 0}_{k,l}(\mathbf{y}) + \varepsilon a_{pqkl}(\mathbf{y})Z^{\alpha\beta 1}_{k,l}(\mathbf{y}) \qquad (4)$$

Let us stress the fact that the position $y_3 = h^{\alpha\beta}$ of the neutral plane of the homogenized plate may not coincide with the middle surface of the homogenized plate. For given $\alpha\beta$, $h^{\alpha\beta}$ is calculated as follows [10]:

$$h^{\alpha\beta} = -A^{\alpha\beta 1}_{\alpha\beta\alpha\beta} / A^{\alpha\beta 0}_{\alpha\beta\alpha\beta}, \qquad (5)$$

where $A^{\alpha\beta\nu}_{ijkl}$ are the rigidities of the plate given by the following formulas [4, 5]:

$$A^{\nu+\mu}_{\gamma\delta\alpha\beta} = \frac{(-1)^{\nu+\mu}}{|P_2|} \int_P a_{\gamma\delta kl}(\mathbf{y})N^{\alpha\beta\nu}_{k,l}(\mathbf{y})y^{\mu}_3 d\mathbf{y} = \frac{(-1)^{\nu+\mu}}{|P_2|} \int_P \sigma^{\alpha\beta\nu}_{\gamma\delta}(\mathbf{y})y^{\mu}_3 d\mathbf{y}.$$

The choice of the neutral plane position following (5) guarantees that in the new coordinate system the out-of-plane rigidity $A^{\alpha\beta 1}_{\alpha\beta\alpha\beta}$ vanishes [10]. Note that the coordinate of the neutral plane $h^{\alpha\beta}$ depends on $\alpha\beta$ (the direction of bending or torsion). Thus, we distinguish several neutral planes in the plate of complex structure.

**3. Numerical analysis of PCPs.** We provide numerical analysis of PCP (3) for various plates containing "layers" of inclusions or holes. In this paper, we perform our computations for two types of inhomogeneous plates: fiber-reinforced plates and plates with systems of channels. The results presented in this paper are representative of other types of plates as well: e.g., plates reinforced by different systems of continuous fibers, filled with discrete particles, plates with tunnels and pores, etc.

*3.1. Plates reinforced by orthogonal layers of fibers.* Consider a plate reinforced by layers of parallel fibers, as displayed in Fig. 1a. This simple reinforcement method is still practically relevant [11, 12]. In our computations, Young's modulus and Poisson's ratio of the fibers and the matrix are $E_f = 170\,\text{GPa}$, $\nu_f = 0.3$ and $E_b = 2\,\text{GPa}$, $\nu_b = 0.36$, correspondingly. These values correspond to a carbon/epoxy composite [13]. In the computations we have $R = 0.45$; $h = 0.1$, and $\delta = 0.1$, see Fig.1. The dimensions of the PC are $h_1 = 1.1$, $h_2 = 3$, $h_3 = 1.1$. The corresponding dimensional values are $\varepsilon R$, $\varepsilon h$, $\varepsilon \delta$, $\varepsilon h_i$ ($i = 1,2,3$); here $\varepsilon$ for carbon fibers varies from 5 to 20 microns [13].

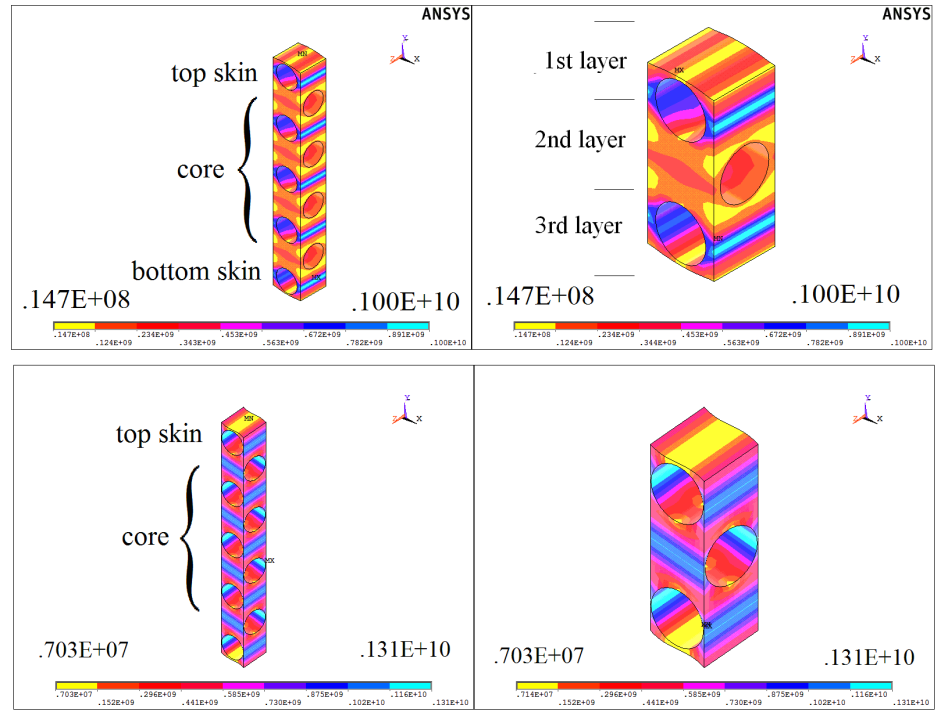

**Fig. 2** *Deformed PCs and the local von Mises stress in the matrix for a 9-layered plate and a 3-layered plate*: *top - in-plane tension; bottom – shift*

The geometry of the PC after the deformation can be observed very well in the figures below. One can receive additional visual information about the stress/strains distribution over the PC by using a proper scalar function of the stress/strains. We take the von Mises stress as such function:

this one is useful for our purposes because it clear indicates the similarity or dissimilarity of the stress/strain states. The von Mises stress also provides some information about the week/strong constitutive elements of the PC.

The deformed PCs and the local von Mises stress are displayed in Fig.2 for a 3-layered PC and 9-layered PC. Only the matrix is displayed. The PCPs for the 3- and 9-layered PCs were solved for the same macroscopic in-plane tension/shift ($v = 0$). The visible matches between the scale bars in Fig.2 indicate a very good coincidence of the local von Mises stress in the 3- and 9-layered PCs.

In Fig. 2 we observe that the local SSS in the top and bottom surfaces of the 9-layered PC (the outer "skin" of the plate) differ from the SSS in the remaining part of the PC (the inner "core"). In the core, the SSS is periodic.

Let us also observe that the SSS in the top/bottom skin of the 9-layered PC are similar to the SSS in the 1st and the 3rd layers of the 3-layered PC. Moreover, the SSS in the core of the 9-layered PC is similar to the 2nd layer of the 3-layered PC.

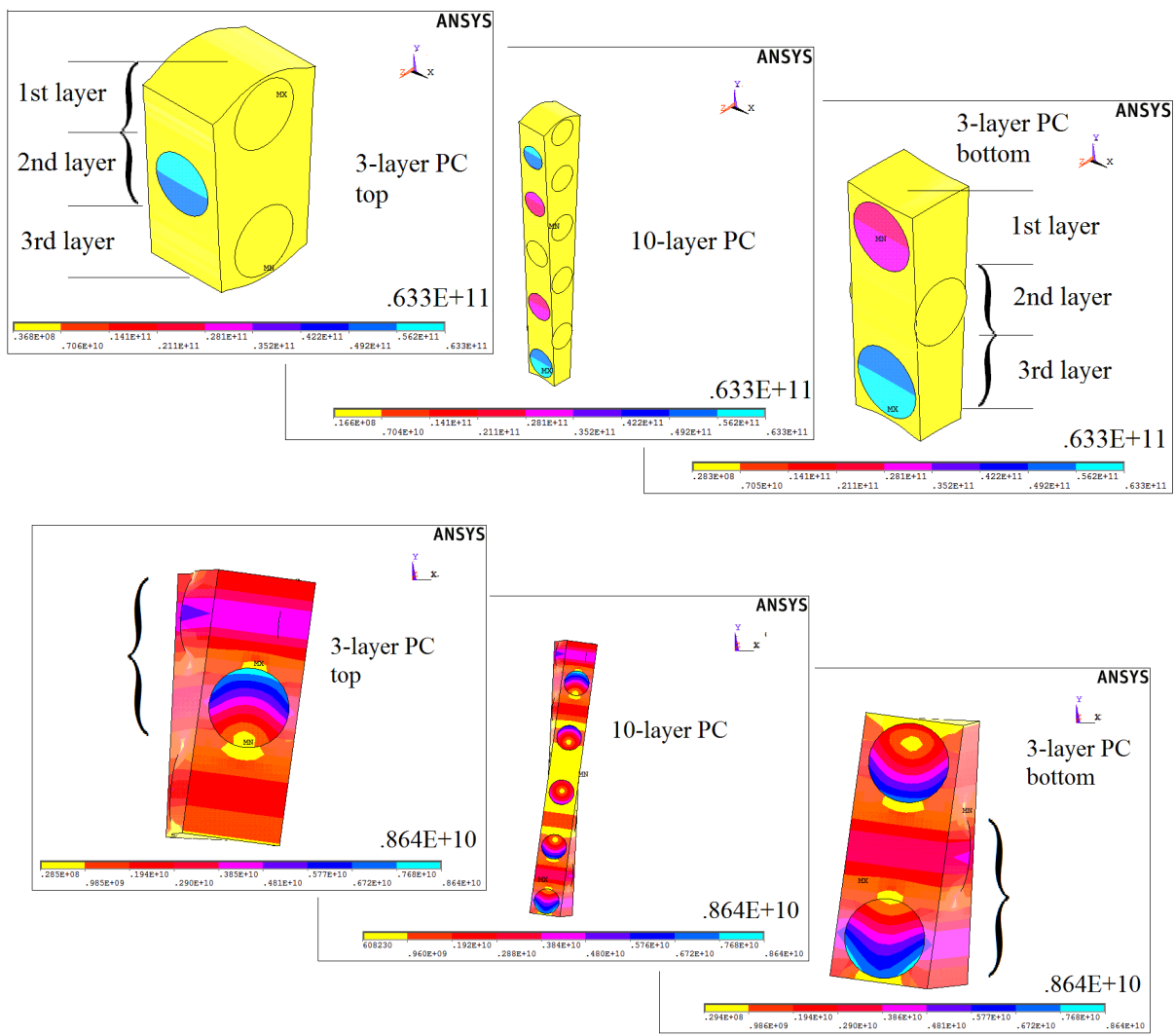

**Fig. 3** *Local von Mises stress in a 10-layered PC and in the top and bottom layers of a 3-layered representative Ps (top – bending, bottom – torsion). Informative layers are marked by brackets*

In Fig. 3 (top) one can observe a pair of 3-layered PCs and a 10-layered PC. The 3-layered PC on the left corresponds to the top of the 10-layered PC. The 3-layered PC on the right corresponds to the bottom of the 10-layered PC. The same description holds for Fig. 3 (bottom). The coincidence between the SSS in the 1st and 2nd layers of the left 3-layered PC and the SSS in the top of the 10-layered PC can be easily observed. There is also a coincidence between the SSS in the 2nd and 3rd layers of the right 3-layered PC and the SSS in the bottom of the 10-layered PC. The visible matches between the scale bars indicate a very good agreement of the values calculated for the respective 3- and 10-layered PCs.

There is no similarity between the SSS in the 3rd layer in the left 3-layered PC and the SSS in the 10-layered PC. There is no similarity either between the SSS in the 1st layer in the right 3-layered PC and the 10-layered PC.

*3.2. Plates with unidirectional systems of holes.* Here we solve a few PCPs for a plate with a system of cylindrical channels parallel to $Ox_1$-axis. The results of numerical computations are given in Fig. 4. The top/bottom fragments of the PCs are zoomed. One can observe that in this case the boundary layer thickness is less than one single fiber.

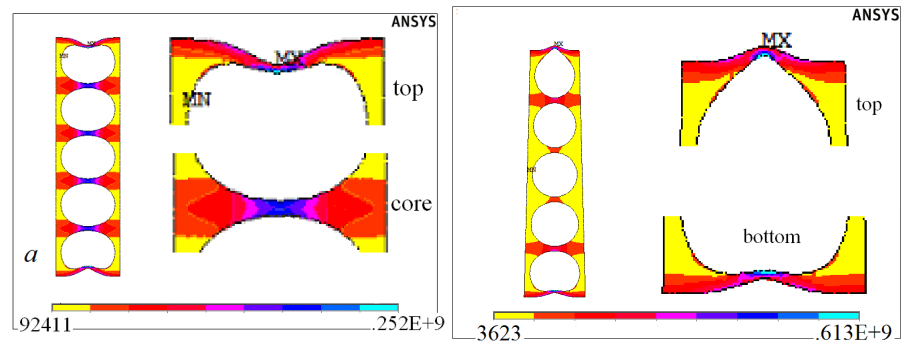

**Fig. 4** *Deformed PCs and the local von Mises stress in the matrix of a 5-layered plate with tunnel cuts: a - in-plane tension; b – bending*

Analyzing the solutions to the PCPs, we conclude that it is impossible to predict where the maximum stress in the inhomogeneous plate occurs without detailed numerical computations.

In this section, we presented solutions to PCP (3) for a fiber-reinforced plate and a plate weakened by layers of channels. We consider a variety of inhomogeneous plates and find out that Fig. 2 – Fig. 4 demonstrate all the typical features of inhomogeneous plates. Let us summarize our findings briefly in the next section.

**4. The top/bottom boundary layers. The "skin" and the "core" of a plate.** The boundary layers at the top/bottom surfaces do not occur in every plate. The known exact solutions to PCP [10] demonstrate that boundary layers never occur at the top/bottom surfaces of homogeneous plates or plates made of homogeneous layers. This firmly establishes that the boundary layers investigated in the present work are fundamentally different from the classical edge layers that are well-known in homogeneous plates [6] and laminated plates [7, 8].

It follows from our computations that the boundary layer thickness is less than that of a single structural layer $S$, see Fig. 1. Consequently, the plate naturally decomposes into three zones: the top layer, the bottom layer, and the remaining layers. We call the top and bottom layers the "skin" of the plate, and the remaining layers the "core" of the plate. The SSS in the skin of a plate and the SSS in its core are substantially different.

**5. A representative model of a multilayered plate**. Usually, the characteristic diameter of the fibers used in composite materials ranges up from a few microns [13]. Thus, the PC of a thin plate may contain several hundreds of fiber layers. This represents a serious obstacle to solving the respective PCP with brute force, even with modern computers. This problem may be solved by introducing the notion of a *representative plate*. Let us also mention that in solid composites the PC coincides with the structural element of the composite, see Fig. 1*c*, and thus such difficulties never appear.

The concept of representative elements is known in the theory of composites since a long time [14, 15]. The concept of a representative plate is not a modification of the latter but rather a new notion in the mechanics of solids. The classical notion of a representative element [14, 15] is based on the "similarity of a portion and the whole". Our concept of representative plate is based on the following two requirements:

1. The SSS in the representative plate is similar to the SSS in the original plate *in some parts*.

2. *For any part* of the original plate there exists a corresponding part in the representative plate. These requirements do not assume that the representative plate is a portion of the original plate.

As follows from above, a multilayered plate PC is separated into three zones: we call them the top skin, the core, and the bottom skin. Since boundary layer thickness is less than one structural layer thickness, then one layer is sufficient to represent every zone. Thus, a 3-layered plate PC can provide us with complete information about the local SSS in a multilayered plate. Such a 3-layered plate is called a *representative plate*.

The use of representative plates makes it possible for accurate computations of the local SSS in an entire plate (its top and bottom skins and its core) with minimal computational resources. To the best of our knowledge, this idea is new.

In a 3-layered representative plate, we distinguish the so-called *informative* and *non-informative* layers. We call a layer *informative* if the SSS in that layer coincides with the SSS in some parts of the original multilayered plate. Else, the layer is *non-informative*. Let us observe that in a 3-layered representative plate two layers are informative, and one layer is non-informative. Note that the non-informative layer still plays an important role in the mechanics of the 3-layered representative plate and thus cannot be removed from consideration.

**6. Wrinkling of the top/bottom surfaces of a plate**. Our computations demonstrate an interesting mechanical phenomenon of wrinkling of the top and bottom surfaces of inhomogeneous plates. The computations above were done for plates with flat upper and lower surfaces. In Fig. 2 – Fig. 4 one can observe that the top and bottom surfaces of the deformed plate are not flat (for in-plane tension) or cylindrical surfaces (for bending). Indeed, they wrinkle up. The latter phenomenon can be very well observed in Fig. 4 for a plate with channel cuts. For a fiber-reinforced plate, one observes wrinkling if the PCP is solved for a twinned PC, see Fig. 5. Here, Fig. 5 corresponds to Fig. 3 (top) zoomed in, with two PCs joined together side-by-side.

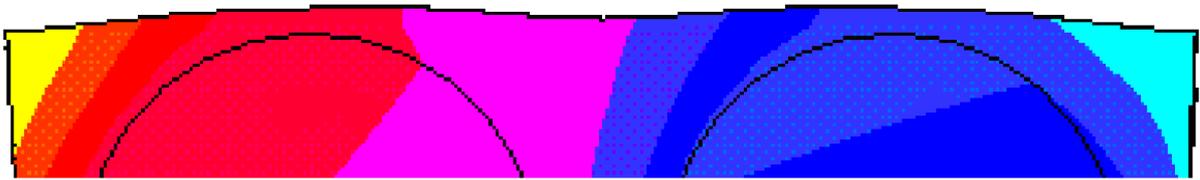

**Fig. 5** *Displacement of the top surface in the twinned PC of a fiber-reinforced plate (bending)*

Although the amplitude of wrinkling has order $\varepsilon \ll 1$, it changes the surface area by a multiple of the order of 1. This may, and will, imply strong surface effects that can prove to be detrimental to the stability or even existence of the whole structure.

**7. Conclusions**. We have found that boundary layers occur at the top and bottom surfaces of inhomogeneous plates. Such boundary layers do not occur in the classical setting and bear no similarity to the classical edge effects on cut surfaces.

The plate naturally decomposes into three zones: its "core" and its top/bottom "skins". Such a 3-layered plate can be representative for any multilayered plate. Thus, one can make conclusions about the strength of a multilayered composite plate by performing the appropriate computations only for the corresponding 3-layered plate.

Surprisingly enough, the thickness of these newly discovered boundary layers is less than that of one structural layer. In practice, such boundary layers manifest themselves thorough wrinkling of the top and bottom surfaces of a given inhomogeneous plate. Although the amplitude of wrinkling is small, it may change the surface are significantly enough to endanger plate's structural integrity. Our approach of representative plates allows to perform numerical modeling and careful study of such possibly adverse effects without extensive numerical computations.

# References

1. Bensoussan A., Lions J.-L., Papanicolaou G. *Asymptotic Analysis for Periodic Structures*. North-Holland, Amsterdam, 1978.
2. Sanchez-Palencia E. *Non-Homogeneous Media and Vibration Theory*. Springer, Berlin, 1980.
3. Bakhvalov N.S., Panasenko G.P. *Homogenization*: *Averaging Processes in Periodic Media*. Kluwer, Dordrecht, 1989.
4. Caillerie D. Thin elastic and periodic plate, *Math. Models Meth. Appl. Sci.*, 1984, 6(1), 159-191.
5. Kohn R.V., Vogelius M. A new model for thin plates with rapidly varying thickness. *Int. J. Solids Struct.*, 1984, 20, 4, 333-350.
6. Van Dyke M. Nineteenth-century roots of the boundary-layer idea. *SIAM review*, 1994, 36(3), 415-424.
7. Pipes R.B., Pagano N.J. Interlaminar stresses in composite laminates under uniform axial extension, *J. Comp. Mater.* 1970, 4(4), 538-548.
8. Herakovich C.T., Post D., Buczek M.B., Czarnek R. Free edge strain concentrations in real composite laminates: experimental-theoretical correlation. *J. Appl. Mech.*, 1985, 52(4), 787-793.
9. Kalamkarov A.L., Kolpakov A.G. *Analysis, Design and Optimization of Composite Structures*. Chichester, Wiley, 1997.
10. Kolpakov A.G. *Stressed Composite Structures*: *Homogenized Models for Thin-Walled Nonhomogeneous Structures with Initial Stresses*. Berlin, Heidelberg, Springer, 2004.
11. Amabili M. *Nonlinear Mechanics of Shells and Plates in Composite, Soft and Biological Materials*. Cambridge, Cambridge University Press, 2018.
12. Kolpakov A.G., Rakin S.I. Homogenized strength criterion for composite reinforced with orthogonal systems of fibers. *Mech. Mater.* 2020, 148, 103489.
13. Agarwal B.D., Broutman L.J., Chandrashekhara K. *Analysis and Performance of Fiber Composites*. 4th Ed., Hoboken, NJ, Wiley, 2017.
14. Dvorak G. *Micromechanics of Composite Materials*. Springer, Dordrecht, 2013.
15. Li Sh., Sitnikova E. *Representative Volume Elements and Unit Cells Concepts, Theory, Applications and Implementation*. Sawston, Woodhead Publishing, 2020.